\documentclass[preprint,12pt]{elsarticle}

\usepackage{amssymb}
\usepackage{amsmath}

\journal{Ultrasonics}

\begin{document}

\begin{frontmatter}



\title{Transforming Breast Cancer Diagnosis: Towards Real-Time Ultrasound to Mammogram Conversion for Cost-Effective Diagnosis}


\author[a]{Sahar Almahfouz Nasser}
\author[b]{Ashutosh Sharma}
\author[a]{Anmol Saraf}
\author[a]{Amruta Mahendra Parulekar}
\author[c]{Purvi Haria}
\author[a]{Amit Sethi}
\affiliation[a]{organization={Electrical Engineering, Indian Institute of Technology Bombay},
            addressline={POWAI}, 
            city={Mumbai},
            postcode={400076}, 
            state={Maharashtra},
            country={India}}
\affiliation[b]{organization={Mechanical Engineering, Indian Institute of Technology Bombay},
            addressline={POWAI}, 
            city={Mumbai},
            postcode={400076}, 
            state={Maharashtra},
            country={India}}
\affiliation[c]{organization={Tata Memorial Hospital},
            addressline={Parel}, 
            city={Mumbai},
            postcode={400012}, 
            state={Maharashtra},
            country={India}}
\begin{abstract}
Ultrasound (US) imaging is better suited for intra-operative settings because it is real-time and more portable than other imaging techniques, such as mammography. However, US images are characterized by lower spatial resolution noise-like artifacts. This research aims to address these limitations by providing surgeons with mammogram-like image quality in real-time from noisy US images. Unlike previous approaches for improving US image quality that aim to reduce artifacts by treating them as `speckle noise', we recognize their value as informative wave interference pattern (WIP). To achieve this, we utilize the Stride software to numerically solve the forward model, generating ultrasound images from mammograms images by solving wave-equations.
Additionally, we leverage the power of domain adaptation to enhance the realism of the simulated ultrasound images. Then, we utilize generative adversarial networks (GANs) to tackle the inverse problem of generating mammogram-quality images from ultrasound images. The resultant images have considerably more discernible details than the original US images.

\end{abstract}

\begin{graphicalabstract}
\begin{figure}[htbp]
  \centering
  \includegraphics[width=\linewidth]{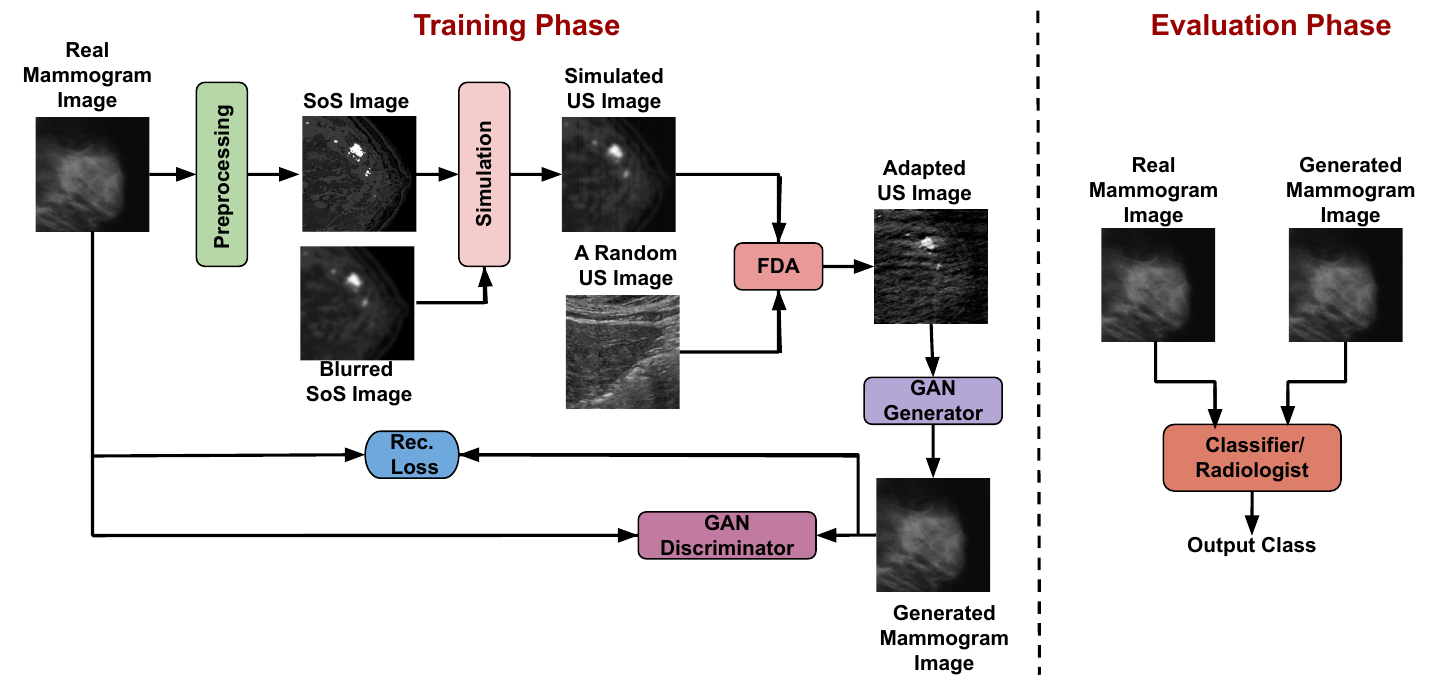}
  \caption{Generating X-ray quality images from ultrasound: We convert an X-ray image (mammogram or CT) into speed of sound, from which we solve wave interference equations to get a simulated ultrasound (US) image. We add high frequency spectral bands from a real US image to make realistic simulated US image. A generative adversarial network then converts a US image into an X-ray image (mammogram or CT) with better visual details than the US image.}
  \label{fig:graphical_abstract}
\end{figure}
\end{graphicalabstract}

\begin{highlights}
\item Novel approach: This research introduces a unique approach to address the limitations of ultrasound imaging by recognizing the value of the wave interference pattern (WIP) in ultrasound images and utilizing it to generate mammogram-like image quality in real-time.
\item Integration of Stride software: The proposed method incorporates the use of Stride software to numerically solve the forward model, allowing for the generation of ultrasound images from mammogram images by solving wave-equations. This integration enables the transformation of ultrasound images into higher quality representations.
\item Domain adaptation and GANs: The study leverages domain adaptation techniques to enhance the realism of simulated ultrasound images. Additionally, generative adversarial networks (GANs) are employed to tackle the inverse problem of generating mammogram-quality images from ultrasound images. These techniques contribute to improving the overall diagnostic potential of the transformed images.
\end{highlights}


\end{frontmatter}


\section{Introduction}
\label{Introduction}

Ultrasound plays a crucial role in image-guided interventions, offering distinct advantages over modalities like MRI, CT, and mammography. These advantages include its portability, real-time imaging capabilities, and lack of bulkiness~\cite{b1}. While MRI, CT, and mammography provide detailed images, they lack real-time capabilities and can be cumbersome~\cite{b44} and~\cite{b45}. However, ultrasound's diagnostic value can be affected by artifacts such as speckle noise, blurring, and shading issues. To improve the diagnostic quality of ultrasound images, we propose a method that involves generating mammogram images from ultrasound. This process requires pairs of mammogram and corresponding ultrasound images to train a deep neural network in a supervised manner. Our approach addresses the forward problem by generating ultrasound from mammogram and the inverse problem by generating mammogram back from the simulated ultrasound using deep learning.

This study presents three main contributions. Firstly, we have developed a novel open-source method for simulating ultrasound images from mammogram scans, departing from the prevalent use of MATLAB or COMSOL in current ultrasound simulation methods. Secondly, we have incorporated a domain adaptation technique to enhance the realism of the simulated ultrasound images. Finally, our proposed method utilizes GANs, which are notoriously challenging to train, to reconstruct mammogram scans from the simulated ultrasound images. Thus, our approach has the potential to provide medical professionals with high-quality mammogram scans derived from ultrasound images, leading to more accurate diagnoses and improved patient outcomes. This advancement represents a significant stride in medical imaging technologies and can contribute to the development of more effective, safe, and precise diagnostic tools.

The subsequent sections of this paper discuss the existing literature on ultrasound simulation, domain adaptation, and image reconstruction techniques. Following that, we will present our proposed method, which includes simulating ultrasound images from mammogram scans, the proposed domain adaptation technique, and the mammogram reconstruction method. We will conclude with a discussion of our results, the limitations of our proposed method, and potential areas for future research and development.

\section{Related Work}

In this section, we will explore various methods for simulating ultrasound, followed by an in-depth examination of domain adaptation techniques. Lastly, we will discuss different generative adversarial networks used for image reconstruction.

\subsection{Ultrasound simulation}

Traditional approaches to denoise ultrasound have considered wave interference as speckle noise \cite{b18}, \cite{b31}, and \cite{b25}. Thus many researchers have primarily focused on simulating speckle noise in ultrasound images, neglecting other interactions between ultrasound and tissue.

Different distribution models have been employed by researchers to approximate speckle noise in simulation, depending on the specific problem being addressed. For example, Goodman utilized a Rayleigh distribution to simulate speckle noise in laser images \cite{b17}, while Wagner et al. employed a Rician model for modeling speckle noise~\cite{b18}. Shankar et al. introduced the Nakagami distribution as a representation of the backscattered ultrasonic echo from the tissue~\cite{b21}. Their method can capture the statistical characteristics of the envelope of the backscattered echo to some extent, encompassing an ensemble of scatterers that exhibit varying number densities, varying cross sections, and the presence or absence of regularly spaced scatterers. In the case of synthetic aperture radar (SAR) images, the Gamma distribution is commonly used as an approximation for speckle noise \cite{b22}. Zimmer et al., in their work on ultrasound liver images, utilized a log-normal distribution to model speckle noise \cite{b24}. Tao et al. demonstrated that the Gamma and Weibull distributions offer more accurate approximations for speckle noise in clinical cardiac ultrasound images compared to normal or log-normal distributions \cite{b15}.
Furthermore, two methods proposed in \cite{b12} and \cite{b13} aim to convert multiplicative noise into additive noise through logarithmic transformation of the image.

Several papers have explored the simulation of ultrasound images from other types of medical images, such as CT and MRI. For instance, Shams et al. \cite{b26} developed an innovative method for simulating ultrasound images from 3D CT scans. Their approach involved edge detection of the CT image to calculate reflection coefficients, and they utilized Field II \cite{b27} to generate the scattering image by randomly placing scatterers with strengths drawn from a normal distribution. However, the authors acknowledged that this method, while effective in creating realistic speckle patterns, is computationally intensive. Simulating a B-mode image with 128 RF scan lines using this technique requires approximately two days. Another method proposed by Kutter et al. \cite{b28} is a real-time simulation-based registration pipeline for ultrasound to CT image alignment. They employed a ray-based model to simulate ultrasound images with the assistance of OpenGL software \cite{b29} and a Lambertian scattering model to generate the scattered signal. The scattering image was created using Field II software. In another study \cite{b30}, the authors adjusted the ultrasound intensity at the probe location, computed the amount of intensity transmitted, reflected, or absorbed for each pixel along each column of the image, and subtracted the reflected and absorbed components from the incident intensity at each pixel. They also introduced post-processing artifacts such as speckle noise and blurring to the ultrasound images, with speckle noise characterized by a Rayleigh distribution. Feng Gu et al. proposed the use of a generative adversarial network (GAN) to model speckle noise in synthetic aperture radar (SAR) images \cite{b11}.

In this work, we present a novel method for simulating the interaction pattern between ultrasound waves and the breast tissue, inspired by the underlying physics of ultrasound image generation, starting from mammogram images of the breast.

\subsection{Domain adaptation}

Domain adaptation techniques aim to improve the performance of machine learning models when there is a mismatch between the distributions of the training data and the test data. Several approaches have been proposed to address this challenge, including transfer learning, domain-specific feature extraction, and domain adversarial training.

Transfer learning methods, such as  pre-training followed by fine-tuning, leverage knowledge acquired from a source domain to enhance performance on a target domain \cite{b40}. Domain-specific feature extraction techniques, such as subspace alignment and kernel mean matching, focus on aligning the distributions of the source and target domains in the feature space \cite{b41}. Domain adversarial training involves training a domain classifier that encourages the feature extractor to produce domain-invariant features \cite{b42}. Other domain adaptation methods include moment matching, maximum mean discrepancy, and co-training, all aimed at reducing the domain shift between the source and target domains \cite{b43}. Fourier-based domain adaptation techniques utilize the Fourier transform to align the source and target domains in the frequency domain, effectively reducing domain shift and improving the performance of machine learning models \cite{b32}. Recent studies have also explored the combination of domain adaptation with other techniques, such as data augmentation and generative models. 

In our work, we aimed to narrow the gap between the distributions of simulated and real ultrasound images by replacing the high frequencies in the Fourier spectrum of simulated images with those from real ultrasound images. This approach preserved the semantic information while achieving more realistic images.

\subsection{Image reconstruction}

In recent years, generative adversarial networks (GANs) have gained significant popularity for their ability to generate high-quality images and other types of data~\cite{goodfellow2014generative}. Various GAN variants have been introduced to enhance their performance and tackle specific challenges. Conditional GANs (cGANs) enable image generation based on specific input conditions as in~\cite{b34} and~\cite{b37}, while Cycle-Consistent GANs (CycleGANs)~\cite{b33} facilitate unpaired image-to-image translation. Several notable GAN architectures have emerged, including Wasserstein GANs (WGANs)~\cite{b35}, which employ a different objective function to enhance training stability, and Progressive GANs (PGANs)~\cite{b36}, which generate images at progressively higher resolutions. Furthermore, methods such as StyleGANs \cite{b38} and BigGANs \cite{b39} have been proposed to improve the diversity and controllability of GAN-generated images.

In our proposed method, we utilize the Pix2pix GAN~\cite{b46} and CycleGAN~\cite{b33} to reconstruct mammogram images from simulated ultrasound images. Further details will be presented in the subsequent sections.

\section{Proposed Method}

Our objective is to generate mammogram-like images from ultrasound images in real-time to provide improved diagnostic capabilities, especially when capturing mammogram images is either infeasible or undesirable due to its associated delays and cost. To accomplish this, we propose a two-stage method. In the first stage, referred to as the forward problem, we begin with mammogram images and generate ultrasound images. We then apply a domain adaptation technique to enhance the realism of the generated ultrasound images. In the second stage, known as the inverse problem, we aim to reconstruct the mammogram images from the generated ultrasound images using GANs. See Figure~\ref{fig:graphical_abstract}  to get an overview of the complete proposed method.


\subsection{Dataset}

The dataset used in this study initially consisted of 3600 mammogram images, with 1200 images for each category: benign, malignant, and normal breast images~\cite{b52}. To prepare the dataset, several preprocessing steps were performed. Firstly, the images were cropped to minimize the presence of dark space. Subsequently, the images were resized uniformly to dimensions of $256 \times 256$ pixels. To ensure the quality and accuracy of the dataset, a manual review of the images was conducted. During this review, any annotations that contained white artifacts (white arrows and/or annotations) were removed from consideration. This process resulted in a reduced dataset, with approximately 1170 images remaining for each category. Furthermore, from the remaining images in each category, a subset of 120 random images was selected. These images were specifically utilized to generate speed of sound images, a specialized type of image representation.

As part of the domain adaptation process, we supplemented our dataset described above by randomly selecting 360 breast ultrasound images from~The Cancer Imaging Archive (TCIA) \cite{b53}. These additional images were chosen to serve as representative samples from the target domain.

Once this paper is accepted, we will release our dataset to the research community, encompassing authentic mammogram images, our simulated ultrasound images, and the actual ultrasound images employed for domain adaptation.

\subsection{Generating speed of sound images from mammogram images}

The initial stage of our proposed method involves converting mammogram images into speed of sound images, which serve as the necessary input for the simulation software to generate ultrasound images corresponding to the mammogram images. To achieve that, we utilize the relationship between the intensity value of a particular pixel in a 2D mammogram image and the corresponding Hounsfield unit (HU) of the underlying tissue to generate speed of sound images. The Hounsfield unit represents the degree of X-ray attenuation in the tissue.
Given a tissue $x$, the HU is given by Equation~\ref{HU}:
\begin{equation}\label{HU}
 HU_x=1000\times \frac{\mu_x-\mu_{water}
 }{\mu_{water}}   
\end{equation}
where $\mu_x$  is the total linear attenuation coefficient of the tissue $x$ at a given x-ray energy. For a tissue $x$ the attenuation coefficient $\mu _x$ can be computed by multiplying the mass density of a tissue $x$  $(\rho_x)$ by the weighted sum of the mass attenuation coefficients of all the elements which compose the tissue $x$, as shown in Equation~\ref{mu_x}:
\begin{equation}\label{mu_x}
    \mu_x = \rho_x \sum_{i}\left(w_i \times \frac{\mu_i}{\rho_i}\right),
\end{equation}          
where $\frac{\mu_i}{\rho_i}$ is the mass attenuation coefficient of the element (i) in cm$^2$/g.

Check Table 1 in our previous paper~\cite{nasser2023simulating} which 
shows the elemental composition ($w_i$ values) of a few of the body tissues taken from the ITIS database \cite{b47}. 

Given the kilovoltage peak applied to the X-ray tube for capturing mammogram images, and $w_i$ values from~\cite{b47}, we can compute the mass attenuation coefficient of each element in the elemental composition of a certain tissue by using NIST software~\cite{b48}. The attenuation refers to the reduction in beam intensity caused by total losses throughout propagation. According to the power law described in \cite{b4} NIST allows us to compute the mass attenuation coefficients of the tissues, which, in turn, allows us to compute their corresponding HUs by substituting the values in Equation~\ref{HU}.

Now, having the speed of sound values and the corresponding HU values of the tissues at 37 Celsius, we can generate the speed of sound images from the corresponding mammogram images. 

We simulate the attenuation of US waves in tissues by Equation~\ref{Attenuation},
\begin{equation} \label{Attenuation}
    U(\alpha_{ref},d)=U_{in} \times e^{\alpha_{ref}d},
\end{equation} 
where $\alpha_{ref}$  is an attenuation parameter related to the properties of the propagation medium taken from \cite{b48}, and $d$ is the propagation depth.



\subsection{Generating US images from speed of sound images}

During the second phase of our proposed simulation approach, we utilize Stride software \cite{b14} to produce ultrasound images from the speed of sound images. Stride is a dedicated open-source library developed for ultrasound computed tomography purposes. Unlike computationally demanding full-waveform inversion methods, Stride is a resource-efficient and user-friendly tool that can be seamlessly run on both CPUs and GPUs. Stride operates on the wave-equation and relies on a domain-specific language called Devito to generate solvers~\cite{b49}.

\begin{figure}[!t]
\centerline{\includegraphics[width=1.0\columnwidth]{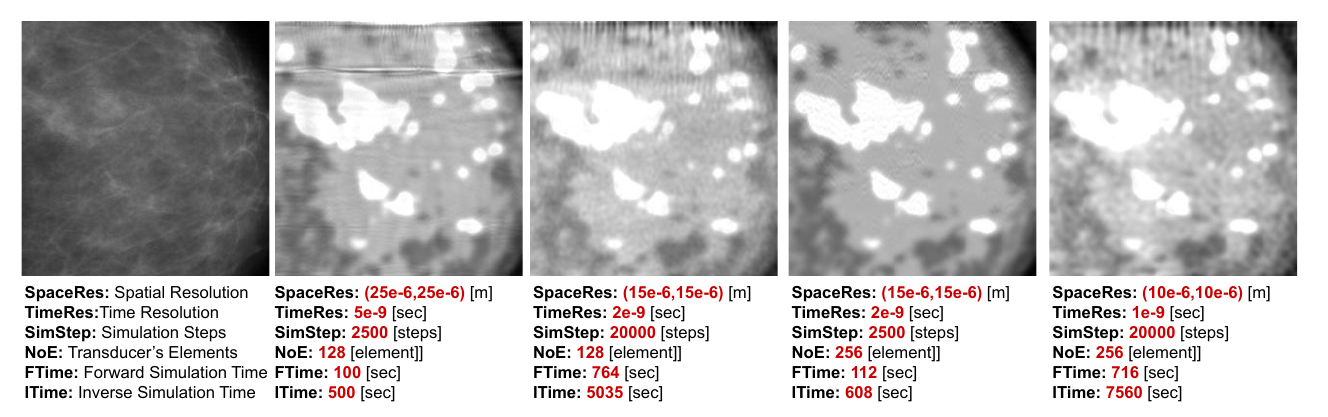}}
\caption{The impact of the simulation settings on the resulting output: The initial image represents the input mammogram, while the subsequent images depict the simulated ultrasound images at various configurations. For all the experiments a linear transducer with a central frequency of 10MHz was simulated.}
\label{simulation_out}
\end{figure}

We input the speed of sound image and adjust various transducer parameters, such as transducer type (linear or curvilinear), number of elements, central frequency, forward simulation duration, time step, and more, into the Stride software as demonstrated in Figure~\ref{simulation_out}. The simulation can be divided into two phases: the forward phase and the inverse phase.

During the forward phase, acoustic waves are generated by sensors and propagate through the medium, interacting with the tissues and reflected on the acoustic boundaries. The receivers capture the reflected waves, and the acquired data is utilized to reconstruct physical properties of the medium, including wave speed and density. In the inverse phase, we begin with an initial estimation of the speed of sound, such as a constant value like 1540 m/sec. However, in our case, we start with a blurred version of the speed of sound image to aid the optimizer in converging, as shown in  Figure~\ref{pipeline}. The reconstruction method focuses on minimizing the difference between the recorded measurements and the numerically simulated ultrasound data to generate the ultrasound image.
 
Hence, utilizing a set of pressure wave field measurements $\boldsymbol{u}$ (captured during the forward pass), we can construct a precise model of the discrete wave velocity $\boldsymbol{C}$ (or $\boldsymbol{m} = \frac{1}{c^2}$) (as output of the inverse pass) by treating it as an optimization problem constrained by a partial differential equation. In this formulation, the objective function can be expressed as follows:

\begin{equation}\label{u2v}
    minimize_{\boldsymbol{m}} \Phi_s(\boldsymbol{m}) = \frac{1}{2}||\boldsymbol{p_ru}-\boldsymbol{d}||_2^2,
    \end{equation}
with $\boldsymbol{u}=\boldsymbol{A}(m)^{-1}\boldsymbol{P}_s^t\boldsymbol{q}_s$, where $\boldsymbol{p_r}$ is the sampling operator of receiver locations, 
$\boldsymbol{P}_s^t$ represents the injection operator at source locations, $\boldsymbol{A}(\boldsymbol{m})$ is the discrete isotropic wave equation matrix, $\boldsymbol{u}$ is the discrete pressure wave field, $\boldsymbol{q}_s$ is the pressure source, and
$\boldsymbol{d}$ is the measured data.

By solving the optimization problem based on the gradient method \cite{b8} \cite{b9} we get:
\begin{equation}
    \delta \Phi_s(\boldsymbol{m}) = \sum _{t=1}^{n_t} \boldsymbol{u}[\boldsymbol{t}]\boldsymbol{\nu}_{tt}[\boldsymbol{t}] = \boldsymbol{J}^T\delta \boldsymbol{d}_s,
\end{equation}
where $n_t$ is the number of steps, $\delta \boldsymbol{d}_s = \boldsymbol{p_ru}-\boldsymbol{d}$ is the data residual between the measured and the modeled data, $\boldsymbol{J}$ is the Jacobian operator, $\boldsymbol{\nu}_{tt}$ is the second-order time derivative of the adjoint wave field $\boldsymbol{A}^T(\boldsymbol{m})\boldsymbol{\nu}=\boldsymbol{P_r}^T\delta\boldsymbol{d}_s$.

Figure~\ref{simulation_out} shows the output of Stride at different simulation settings for the same input mammogram image.

\subsection{Simulated ultrasound to realistic ultrasound conversion}

The results obtained from the Stride software indicated that it was unable to simulate the speckle noise present in real ultrasound images. This discrepancy caused a significant difference in the distribution of the simulated images compared to that of real ultrasound images. To address this limitation of Stride and to align the distributions more closely, we developed a domain adaptation method. In this section, we will provide a detailed explanation of our proposed domain adaptation approach.

Taking inspiration from the approach presented in \cite{b32}, which introduced a novel method for domain adaptation in semantic segmentation tasks utilizing Fourier domain analysis, our work aims to enhance the fidelity of simulated ultrasound images by closely resembling real ultrasound images. The referenced authors sought to improve the performance of semantic segmentation models across diverse domains, such as varying lighting conditions or distinct datasets. Their proposed method involves two main steps: First, both the source and target domain images undergo transformation into the Fourier domain. Subsequently, a domain adaptation network is trained to map the source domain images to the target domain images within the Fourier domain. This network is then employed to adapt the semantic segmentation model, previously trained on the source domain, to perform effectively on the target domain. Notably, the authors demonstrated successful adaptation even across substantially different domains, such as photographs taken during distinct seasons and times of day.

\begin{figure}[!t]
\centerline{\includegraphics[width=1.0\columnwidth]{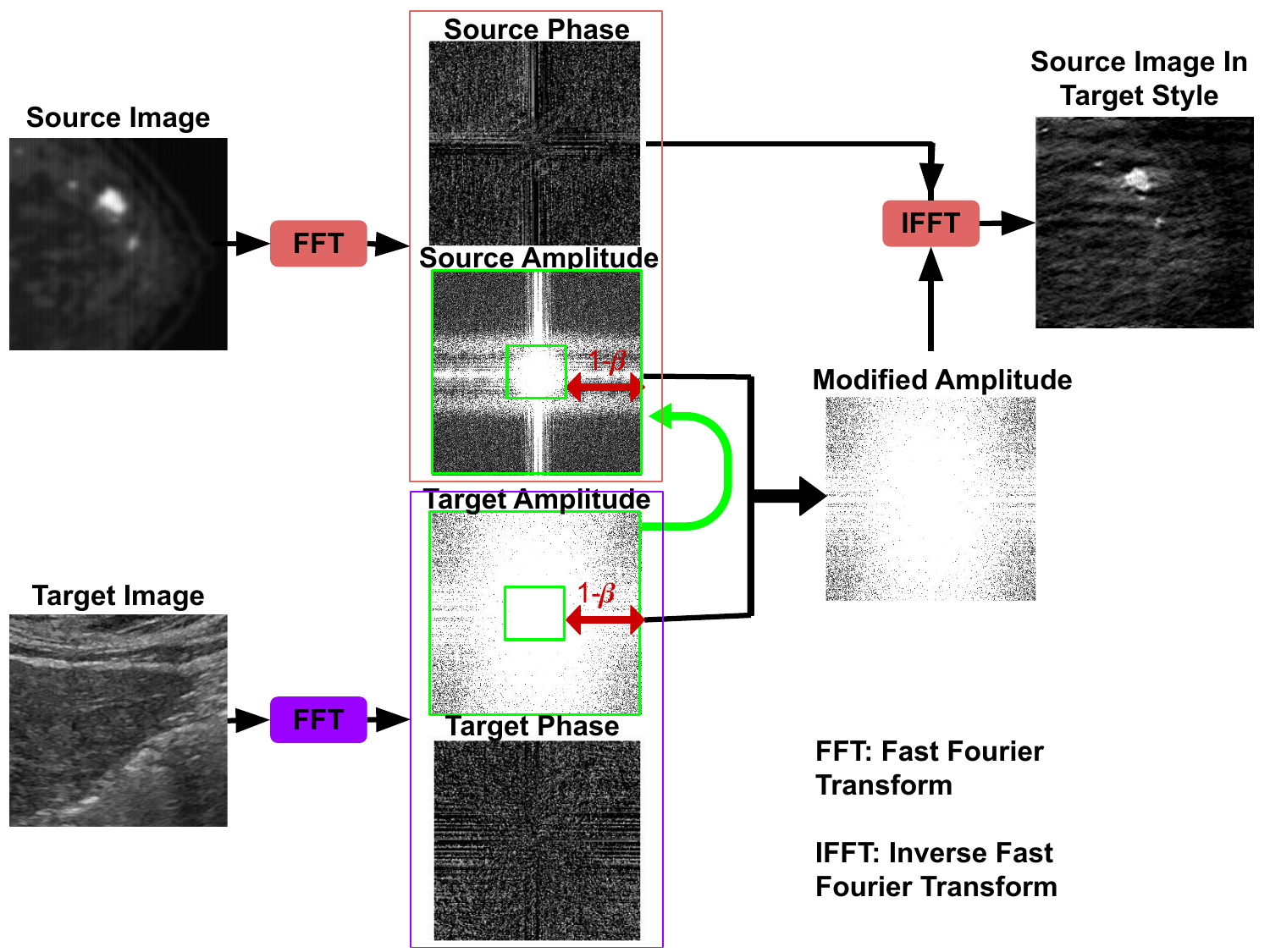}}
\caption{Spectral transfer involves the process of mapping a source image to a target style while preserving its semantic content. To achieve this, the high-frequency component of the source image's spectrum is replaced with that of a randomly selected target image, effectively transferring its style. This technique results in a perceptually smaller domain gap and enhances the quality of the generated of mammogram images from simulated ultrasound images.}
\label{fda_diagram}
\end{figure}

Building upon this concept and considering our specific objective of generating simulated ultrasound images closely resembling real ultrasound images, we employed a similar technique employing Fourier transformation, as show in Figure~\ref{fda_diagram}. Specifically, we utilized a mapping process in the Fourier domain to align the source domain images (simulated ultrasound images) with the target domain images (real ultrasound images). While the approach presented in the referenced paper aimed to modify color and lighting to match the target image, our focus differed. We sought to add noise to the simulated image, thereby bringing it closer to the characteristics of real ultrasound images. Consequently, we replaced the high frequencies of the Fourier spectrum of the simulated image with the high frequencies from the Fourier spectrum of the real ultrasound image to bring the noise texture, but we retained the low frequency component to retain the overall style of the image.

\begin{figure}[!t]
\centerline{\includegraphics[width=1.0\columnwidth]{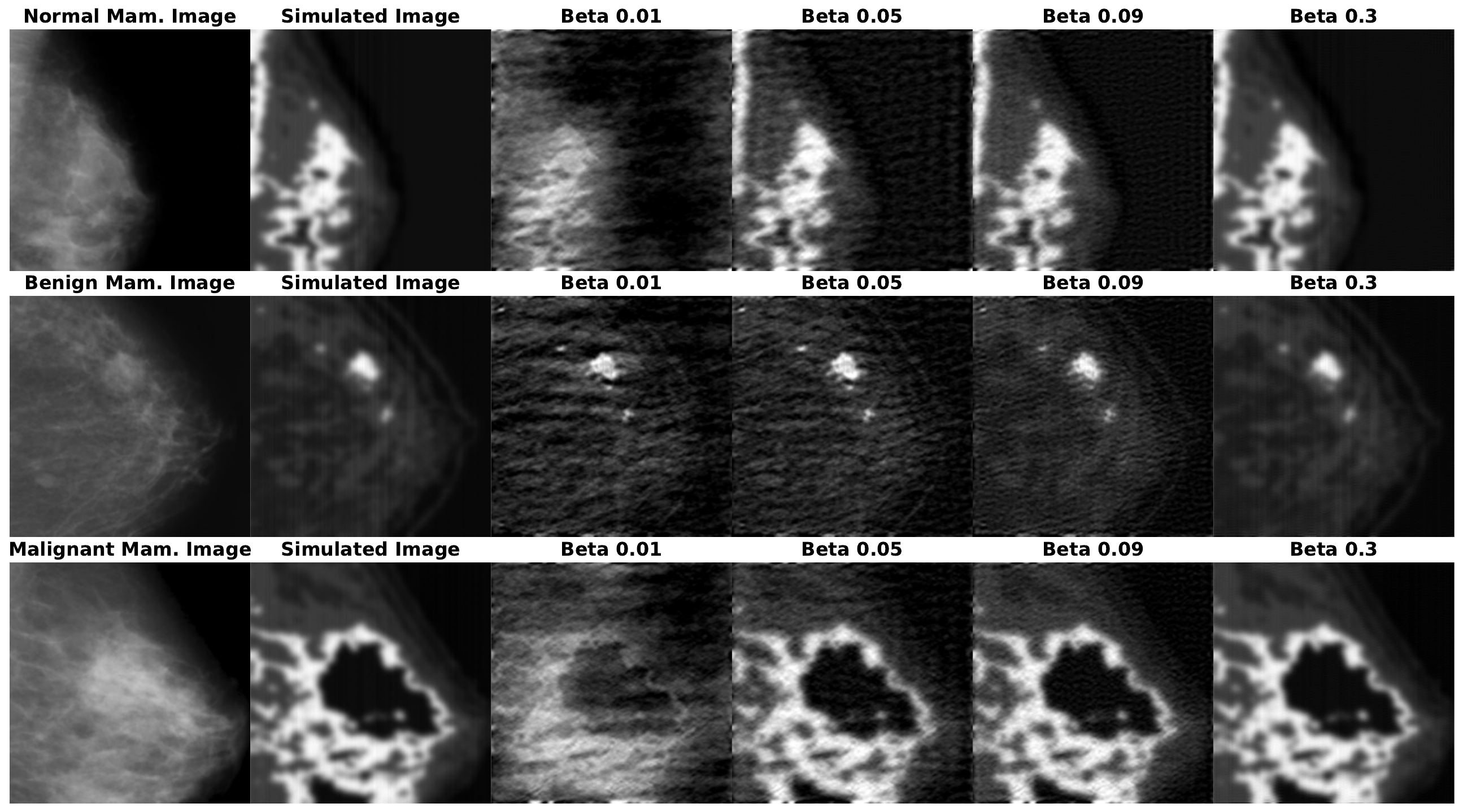}}
\caption{Effect of domain adaptation on US generation: The three examples -- normal, benign, and malignant -- are arranged in three rows from top to bottom. In each row, the columns correspond to the input mammogram, the ultrasound image simulated by Stride software, and the outputs of our method for domain adaptation at betas 0.01, 0.05, 0.09, and 0.3, represented by columns 3 to 6, respectively.}
\label{fda_images}
\end{figure}

Figure~\ref{fda_images} showcases three examples that belong to distinct classes of mammogram images. Additionally, it displays their corresponding simulated ultrasound images as well as the adapted ultrasound images computed at different values of Beta.

\begin{figure}[!t]
\centerline{\includegraphics[width=1.0\columnwidth]{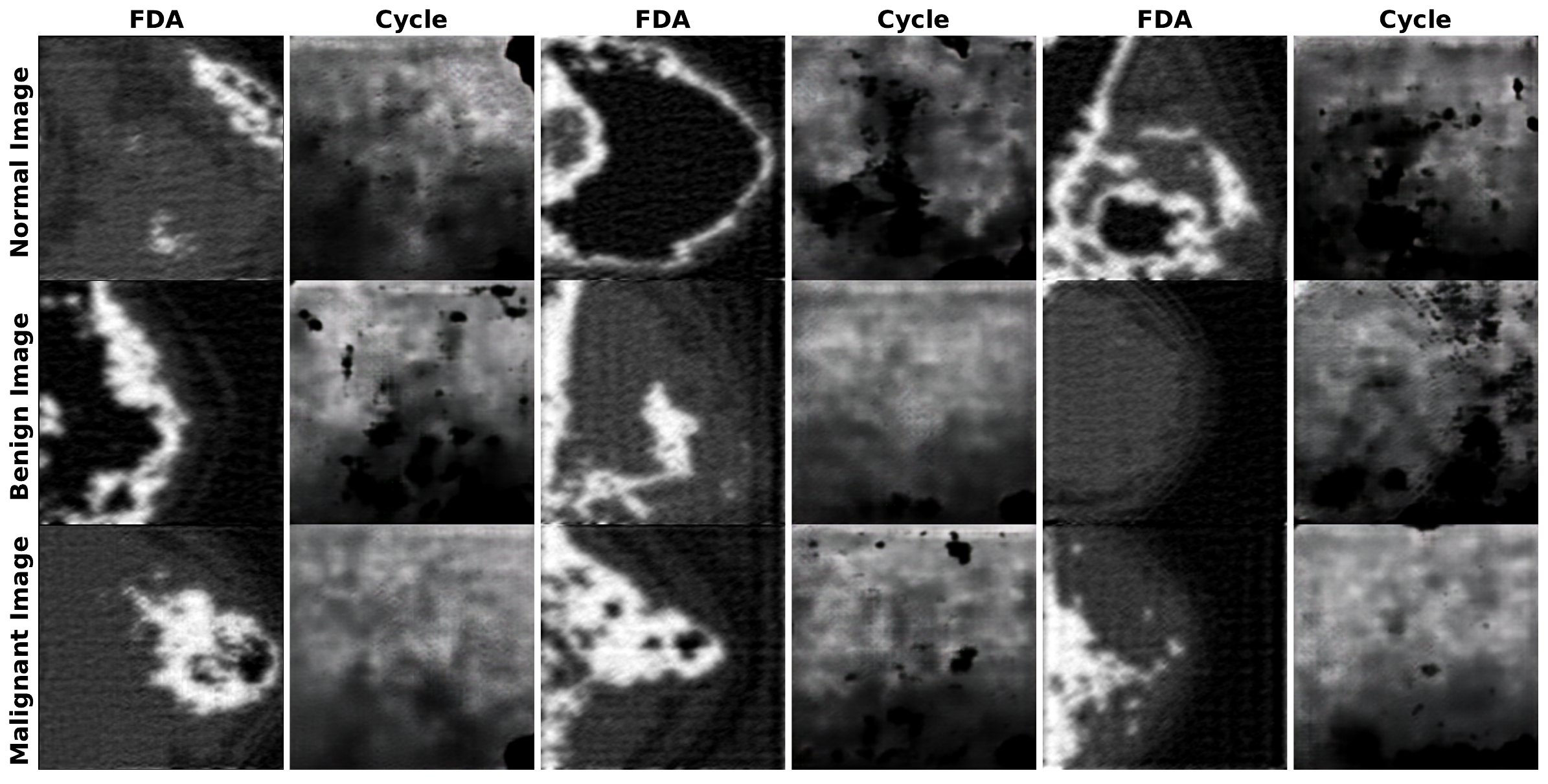}}
\caption{A comparative analysis of image quality between FDA-generated images and cycleGAN-generated images.}
\label{fda_cycle}
\end{figure}

In our supplementary domain adaptation experiment, we opted to employ cycleGAN instead of our modified FDA method to generate real US images from simulated US images. However, this alternative approach revealed several drawbacks. Primarily, as GANs have a tendency to hallucinate data, the semantic information that was preserved using FDA could no longer be maintained. Additionally, as illustrated in Figure~\ref{fda_cycle}, the generated images exhibited significant issues with artifacts, greatly compromising their quality.

\subsection{Mammogram reconstruction}

In the inverse problem, we employ a supervised approach to train a GAN that reconstructs the mammogram image back from the simulated ultrasound image.

Our approach to image-to-image translation, is based on the method presented in the paper referenced as~\cite{b33}. In~\cite{b33} the objective of the authors was to generate realistic images in one domain based on images from another domain. For example, they aimed to convert horse images into zebra images or transform maps into satellite images. The method consists of two primary components: a generator network and a discriminator network. The generator network learns to convert images from one domain to another, while the discriminator network learns to differentiate between real and generated images. The training process involves an adversarial approach to enhance the quality of the generated images.
To ensure the consistency between the generated and original images, a cycle-consistency loss was introduced. This function penalizes the generator network for producing images that cannot be reconstructed back to the original domain. By incorporating this loss function, the generator is discouraged from generating unrealistic or inconsistent images.

In our specific case of generating mammogram-like images from simulated ultrasound images in the target domain, we applied the Pix2Pix GAN method described in Zhu et al.'s work \cite{b33}.

In our experiments, the discriminator loss represents the adversarial loss, whereas the generator loss is a weighted combination of the adversarial loss, the L1 loss (to preserve style information), and the perceptual, Laplacian, or wavelet loss (to preserve semantic information) as demonstrated in~\cite{b37}.
The objective function is given by Equation~\ref{ad_loss}.
\begin{equation}\label{ad_loss}
   L = \underset{G}{min} \underset{D}{max} \quad \mathbb{E}_{x\sim p_{data}(x)} [logD(x)] + \mathbb{E}_{z\sim_{z (z)}}[log(1-D(G(z))].
\end{equation}
The generator loss represented by Equation~\ref{g_loss}
\begin{equation}\label{g_loss}
    L_{G} = \alpha_1 L_P+\alpha_2 L_1 + \alpha_3 L_{ad},
\end{equation}
where $L_P$ represents the perceptual loss, and $L_{ad}$ represents the adversarial loss. From our experiments we found that the best values for the hyperparameters are $\alpha_1=0.0$, $\alpha_2=10.0$, and $\alpha_3 = 1.0$. In certain experiments, we replaced the perceptual loss with the Laplacian loss and the wavelet, as indicated in Table~\ref{table:RecResults}. The Laplacian loss can be defined as the L1 loss between the Laplacian images of the reconstructed mammogram image and the corresponding ground truth mammogram image. The wavelet loss is calculated as the mean L1 loss between the corresponding channels of the wavelet-transformed reconstructed mammogram image and the corresponding wavelet-transformed ground truth mammogram image. We observed that incorporating wavelet loss leads to the generation of images with enhanced fine details compared to using perceptual loss. However, this improvement comes at the cost of an increased presence of artifacts (black regions) in the generated images.

\begin{figure}[!t]
\centerline{\includegraphics[width=1.0\columnwidth]{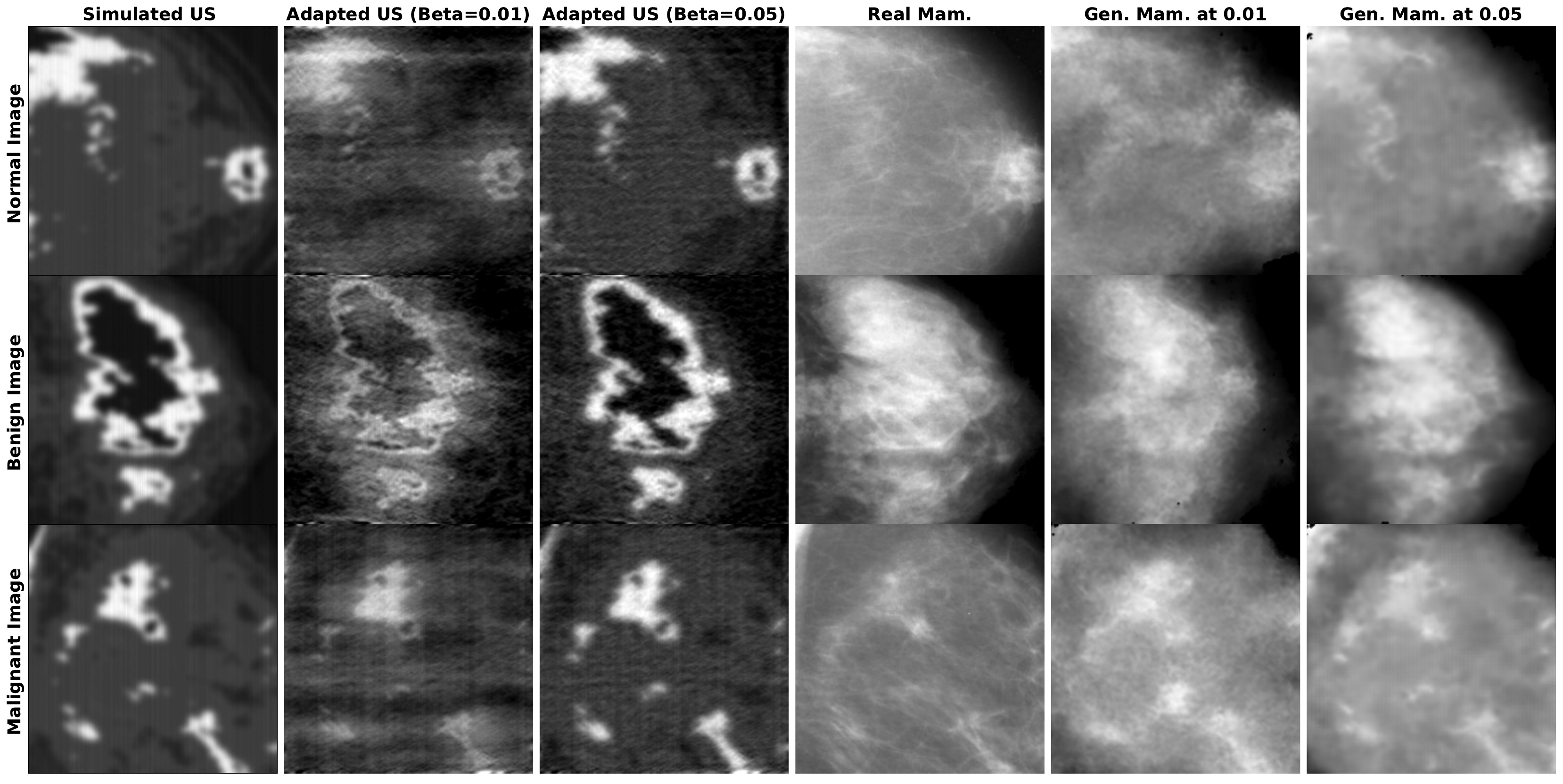}}
\caption{The testing results of the GAN for the three classes normal, benign, and malignant are displayed in rows from top to bottom in this figure. The GAN was fed with ultrasound images at Beta = 0.01 and Beta = 0.05 as inputs.}
\label{gan_01}
\end{figure}

The performance of the GAN on the testing data is depicted in Figure~\ref{gan_01}, highlighting three distinct examples from the testing dataset: normal, benign, and malignant.

The evaluation metrics for the testing dataset are presented in Table~\ref{table:RecResults}. We assessed the mean square error (MSE), peak signal to noise ratio (PSNR), and structural similarity (SSIM) between the reconstructed mammogram image and the corresponding ground truth mammogram image. Two different construction methods, namely Pix2Pix and CycleGAN, were employed using various values of $\beta$ and different combinations of loss functions. From the table, it is evident that Pix2Pix outperforms CycleGAN when utilizing the base loss and at a $\beta$ value of 0.05. In an attempt to enhance the structural information in the reconstructed mammogram image, we experimented with alternative losses like perceptual loss, Laplacian loss, and wavelet loss; however, none of them yielded improved results.

\renewcommand{\arraystretch}{1.5}
\begin{table*}[t]
\caption{This table presents the evaluation metrics for various experiments conducted, including mean square error (MSE), peak signal-to-noise ratio (PSNR), and structural similarity (SSIM). The experiments are differentiated by the types of losses employed, with "PL" representing Perceptual loss, "LL" indicating Laplacian loss, and "W" representing wavelet loss. "Rec" refers to the experiment in which we kept the weight of the reconstruction loss ten times the weights of other losses.}
\centering
\scalebox{0.9}{
\begin{tabular}{|l|l|l|l|l|}
 \hline
 \textbf{Experiment} & \textbf{MSE} & \textbf{PSNR} & \textbf{SSIM} \\ 
 \hline\hline
  Pix2pix, $\beta=0.01$ &	0.014 $\pm$ 0.008 &18.54 $\pm$ 20.97 & 0.806 $\pm$ 0.071\\
  \textbf{Pix2pix, $\beta=0.05$}& \textbf{0.008 $\pm$ 0.008}& \textbf{20.97$\pm$20.97} &\textbf{0.870 $\pm$ 0.070}\\
  Pix2pix, $\beta=0.01$, PL& 0.018 $\pm$ 0.012& 17.45 $ \pm$ 19.21 &0.784 $\pm$ 0.064\\
  Pix2pix, $\beta=0.05$, PL& 0.010 $\pm$ 0.010& 20.00 $\pm$ 20.00 &0.837 $\pm$ 0.081\\
  Pix2pix, $\beta=0.05$, LL&\textbf{0.008 $\pm$ 0.008}& \textbf{20.97$\pm$20.97} &0.862 $\pm$ 0.077\\
  Pix2pix, $\beta=0.05$, W&\textbf{0.008 $\pm$ 0.008}& \textbf{20.97$\pm$20.97} &0.862 $\pm$ 0.076\\
  CycleGAN, $\beta=0.05$& 0.040 $\pm$ 0.021& 13.98 $\pm$ 16.78 &0.636 $\pm$ 0.113\\
  CycleGAN, $\beta=0.05$, PL& \textbf{0.008 $\pm$ 0.008}& \textbf{20.97$\pm$20.97} &0.856 $\pm$ 0.070\\
  CycleGAN, $\beta=0.05$, Rec& 0.01 $\pm$ 0.01& 20.00 $\pm$ 20.00 &0.785 $\pm$ 0.066\\
 
 \hline
\end{tabular}
}
\label{table:RecResults}

\end{table*}

\subsection{Mammogram reconstruction from real ultrasound images}

To realize our ultimate objective of reconstructing mammogram images from actual ultrasound images, we conducted three experiments. Initially, we curated a dataset comprising unpaired real ultrasound images and real mammogram images, with each pair belonging to the same class—normal, benign, or malignant.
We employed CycleGAN, enabling the generation of mammogram images from real ultrasound images. This involved training from scratch in one experiment and using our pretrained CycleGAN on simulated ultrasound images, which was integrated into two separate experiments.

\begin{figure}[!t]
\centerline{\includegraphics[width=1.0\columnwidth]{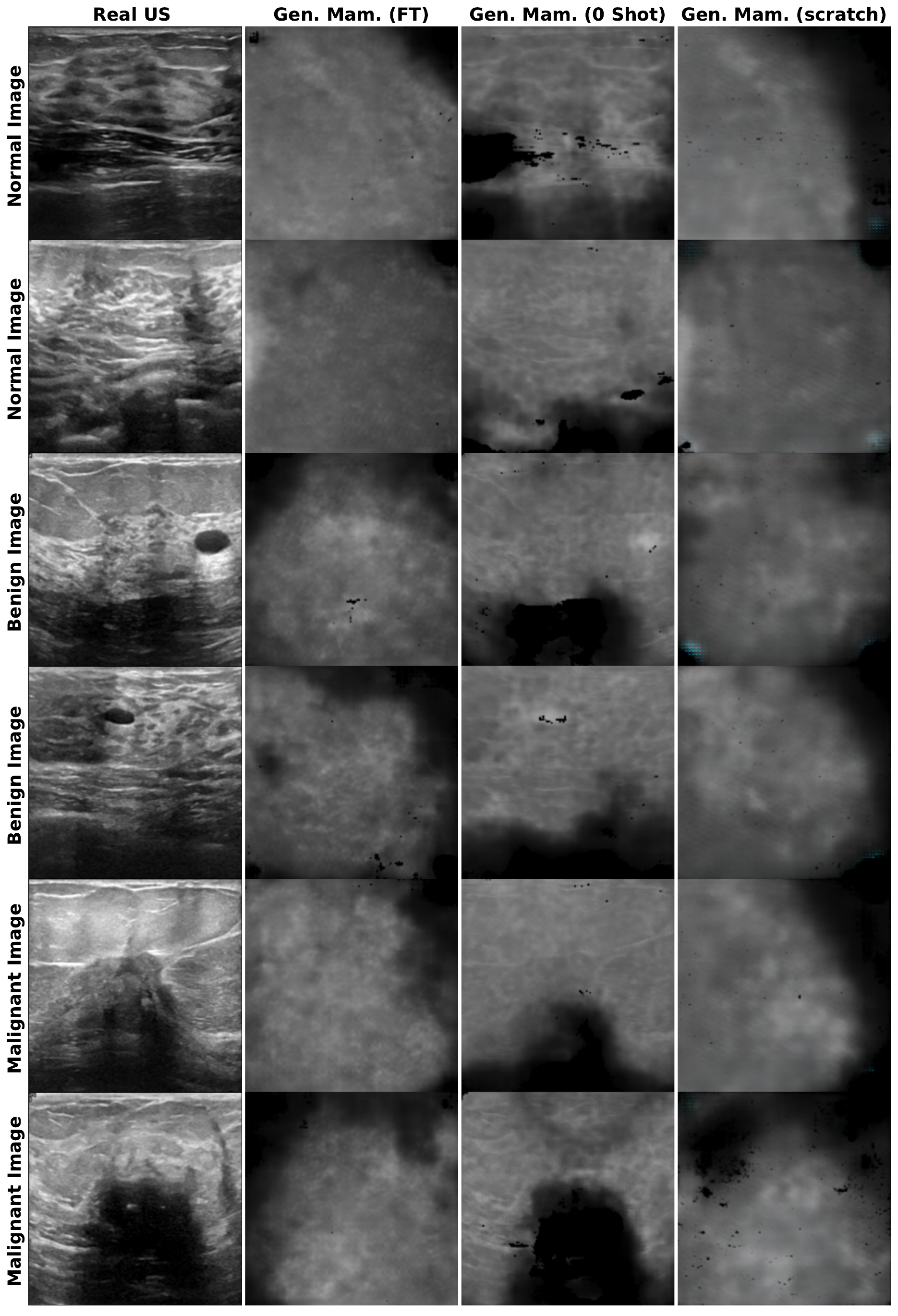}}
\caption{Transforming real ultrasound images into mammogram counterparts. The figure exhibits three rows showcasing the results, including two input images from each class—normal, benign, and malignant—arranged from top to bottom. 'Gen. Mam.' denotes the generated mammogram images, 'FT' represents fine-tuning, '0 Shot' corresponds to the zero-shot approach, and 'Scratch' refers to training from scratch without starting from weights of cycleGAN pretrained on simulated US images. }
\label{real_US}
\end{figure}

In the first experiment, termed 'Zero shot learning', we deployed the pretrained network in testing mode, taking real ultrasound images as input and presenting the corresponding reconstruction results, as depicted in Figure~\ref{real_US}.
In the second experiment, we performed fine-tuning on our pretrained CycleGAN using the new set of unpaired images, referenced as 'FT' in Figure~\ref{real_US}. Encouragingly, Figure~\ref{real_US} demonstrates that the results obtained from the fine-tuning experiments are plausible, igniting hope that this method could potentially replace mammogram imaging—an approach that poses some harm to patients—and offer more precise diagnoses using ultrasound images alone.
Despite being trained on simulated ultrasound and mammogram images showing the entire breast, our model exhibits the remarkable ability to generate localized regions of mammogram images when provided with a real ultrasound image that has a narrower field of view compared to a standard mammogram image unlike the results we got when we trained from scratch.
Furthermore, as illustrated in Figure~\ref{real_US}, when training from scratch, the resulting images appear blurry compared to those generated by the fine-tuned model. In addition to this blurriness, training from scratch introduces undesirable artifacts, such as blue-color artifacts and numerous black patches.

\section{Conclusion and Discussion}
\label{sec:conclusion}

In this study, we have developed an innovative pipeline aimed at transforming ultrasound images into mammogram-like images in real time. The purpose of this pipeline is to enhance the diagnosis of breast cancer in situations where mammography is challenging or not feasible. Through the incorporation of ultrasound simulation, domain adaptation, and mammogram reconstruction techniques, our pipeline has demonstrated notable achievements.
The results obtained from using real US images to generate mammogram images confirmed the success of our pipeline in improving the quality and diagnostic potential of ultrasound images, especially in the context of breast cancer diagnosis. 
To further assess the effectiveness of our approach, we are planning to conduct evaluations on two levels. Firstly, at the machine level, we intend to employ a deep learning classifier to classify the generated mammogram images. Secondly, at the human level, we will seek the expertise of a radiologist to evaluate and classify the generated mammogram images. Moving forward, we are actively engaged in further research to transform our pipeline into user-friendly software that can be readily utilized in clinical settings. This software has the potential to facilitate easier and more cost-effective breast cancer diagnosis.

\section*{Acknowledgement}

The financial support from the Qualcomm Innovation Fellowship Award India, was essential in enabling this work. We are grateful to Mr. Carlos Cueto, the developer of Stride from Imperial College London for making Stride open source and for providing us with his invaluable feedback and assistance.

\section{Declaration of generative AI and AI-assisted technologies in the writing process}

During the preparation of this work the authors used ChatGPT in order to improve the language and readability of the paper. After using this tool, the authors reviewed and edited the content as needed and take full responsibility for the content of the publication.

\bibliographystyle{IEEEtran}
{\small
\bibliography{references}}
\end{document}